\documentclass[showkeys,preprint,showpacs,amsmath,amssymb,aps,prb,floatfix,endfloats*]{revtex4}

\usepackage{graphicx}% Include figure files
\usepackage{longtable}% Include figure files
\usepackage{dcolumn}% Align table columns on decimal point
\usepackage{bm}% bold math

\def\etal{{\em et al. }}
\def\abinit{{\it ab-initio\ }}
\def\Abinit{{\it Ab-initio\ }}
\def\vk{{\bf k }}
\def\rhm{\bar{\rho}}
\def\mjm2{mJ/m$^{2}$ }
\def\al3ti{Al$_3$Ti }
\def\ti3al{Ti$_3$Al }
\def\alphati{$\alpha$-Ti }
\def\gtial{$\gamma$-TiAl }
\def\evm{$E_v^m$ }
\def\evf{$E_v^f$ }
\def\l10{L1$_0$ }

\begin{document}

\title{ Interatomic potentials for atomistic simulations of the Ti-Al system}

\author{Rajendra R. Zope and Y. Mishin} \email{rzope@scs.gmu.edu,
 ymishin@gmu.edu} \affiliation{School of Computational Sciences, George
 Mason University, Fairfax, VA 22030}

\date{\today}


%Use showkeys class option if keyword display desired

\begin{abstract}
Semi-empirical interatomic potentials have been developed for Al, \alphati,
and \gtial within the embedded atom method (EAM) formalism by fitting to a
large database of experimental as well as \abinit data.  The {\it ab
initio}\ calculations were performed by the linearized augmented plane wave
(LAPW) method within the density functional theory to obtain the equations
of state for a number of crystal structures of the Ti-Al system. Some of the
calculated LAPW energies were used for fitting the potentials while others
for examining their quality. The potentials correctly predict the
equilibrium crystal structures of the phases and accurately reproduce their
basic lattice properties. The potentials are applied to calculate the energies
of point defects, surfaces, and planar faults in the equilibrium structures.
Unlike earlier EAM potentials for the Ti-Al system, the proposed potentials
provide a reasonable description of the lattice thermal expansion,
demonstrating their usefulness for molecular dynamics and Monte Carlo
simulations at high temperatures.  The energy along the tetragonal
deformation path (Bain transformation) in \gtial calculated with the EAM
potential is in a fairly good agreement with LAPW calculations.  Equilibrium
point defect concentrations in \gtial are studied using the EAM potential.
It is found that antisite defects strongly dominate over vacancies at all
compositions around stoichiometry, indicating that \gtial is an antisite
disorder compound in agreement with experimental data.
\end{abstract}

\maketitle

\section{\label{sec:intro}Introduction\protect\\ }

In the recent years, intermetallics alloys based on the gamma titanium
aluminide TiAl have been the subject of intense research due to their
potential for applications in the aerospace and automobile
industries.\cite{Kim94,ABW93,YIKMS96,YII00} Such alloys have an excellent
oxidation and corrosion resistance which, combined with good strength
retention ability and low density, make them very advanced high temperature
materials.  A study of fundamental properties such as the nature of
interatomic bonding, stability of crystal structures, elastic properties,
dislocations, grain boundaries, interfaces, as well as point defects and
diffusion are therefore warranted in order to gain more insight into the
behavior of these intermetallic alloys under high temperatures and
mechanical loads.  Over the past few years, numerous investigations, both
experimental and theoretical, have been devoted to the study of such
properties.\cite{HNCW02,App01,MGF98,SRD98,ZRD01,FY93,IKNIYK01,MH00,FZY95}
Accurate \abinit studies of the structural stability, elastic properties,
and the nature of interatomic bonding have been reported for \gtial as well
as other stoichiometric alloys of the Ti-Al
system.\cite{HWGFOX91,HWFOX90,Fu90} However, the application of \abinit
methods to atomistic studies of diffusion, deformation, and fracture are
limited due to the prohibitively large computational resources required for
modeling point defects, dislocations, grain boundaries, and fracture cracks.
Such simulations require large simulation cells and computationally
demanding techniques such as molecular dynamics and Monte
Carlo. Semiempirical methods employing model potentials constructed by the
embedded atom method (EAM)\cite{DB83_0,DB84,Daw93} or the equivalent
Finnis-Sinclair (FS) method\cite{FS84} are particularly suitable for this
purpose. These methods provide a way of modeling atomic interactions in
metallic systems in an approximate manner allowing fast simulations of large
systems. Several studies applying these methods to a variety of properties
of \gtial such as planar faults, dislocations, and point defects have been
reported in the literature.\cite{SRD97,PWSV99,MH00} The effectiveness of
semiempirical methods obviously depends upon the quality of the model
potentials employed.  Recent
studies\cite{EA94,KASVE93,LAEM96,LWGVRS98,FOZ99,MFMP99,BAEJ99,MMPVK01,BAS01,MMP02}
have shown that the incorporation of \abinit data during the fitting of
interatomic potentials can significantly enhance their ability to mimic
interatomic interactions. For example, Bakes \etal\cite{BAS01} examined the
range of interatomic forces in aluminum using model potentials and \abinit
methods. They found that potentials that included \abinit data during the
fitting procedure could reproduce \abinit forces much more accurately than
potentials fit to experimental data only.

In the present work we explore the possibility of constructing a reliable
interatomic potential for the Ti-Al system. To this end, we develop
EAM-type interatomic potentials for \gtial and the component elements Ti
and Al by fitting to a large database of experimental properties and \abinit
structural energies of these phases. The \abinit database has been generated
by density functional calculations using the linearized augmented plane
wave (LAPW) method within the generalized gradient approximation (GGA) for
the exchange-correlation effects. The \abinit data are used in the form of
energy-volume relations (equations of state) of various structures of Al,
Ti, and \gtial. The energy along the Bain transformation path between the
L1$_0$ and B2 structures of TiAl has also been calculated in this work.

While many EAM-type potentials have been reported for
Al,\cite{EA94,VC87,HVF99,MFMP99,BAS01} relatively few attempts have been
made to create such potentials for Ti\cite{PS92,BJ94,FMP95} and
TiAl.\cite{Farkas94,RWP91,RWSD95,CYL99,WWJG01,SRD97,PWSV99} Titanium, like
other transition metals, cannot be expected to follow the EAM model as
accurately as noble metals usually do. In contrast, Al is known to lend
itself to the EAM description quite
readily.\cite{EA94,VC87,HVF99,MFMP99,BAS01} The \gtial compound is probably
a borderline case. Some of the previous EAM potentials for \gtial had a
reasonable success in modeling lattice properties and extended lattice
defects.\cite{Farkas94,SRD97,PWSV99} On the other hand, Paidar
\etal\cite{Paidar_TiAl} calculated deformation paths between different
structures of TiAl by an \abinit method and with a Finnis-Sinclair potential
and found only a qualitative agreement between the two calculation
methods. In this work we are trying to gain a better understanding of the
applicability range of the EAM model for \gtial. In particular, we want
explore the possibility of obtaining a high-quality EAM potential for \gtial
by incorporating \abinit data into the fitting database. Our results suggest
that the quality of an EAM potential can indeed be improved this way, and
that the potential proposed here can be useful for atomistic simulations of
\gtial.

\section{\label{sec:EAM}Construction of EAM potentials}

\subsection{\label{sec:datab}Database of fitted physical properties}

We have first constructed EAM potentials for pure Al and Ti followed by
fitting a cross-interaction potential Ti-Al. The database of physical
properties employed in the fitting procedure consisted of two categories.
The first category is comprised of experimental data for the lattice
constant, $c/a$ ratio, cohesive energy, and elastic constants.  For pure Al
and Ti it also included the vacancy formation energy and the linear thermal
expansion factors at several temperatures. The second set of properties
consisted of \abinit energy differences between various crystal structures.
Such differences are necessary to ensure the correct stability of the
experimentally observed ground state structures against other possible
structures and to sample a large area of configuration space away from
equilibrium.

The \abinit database consists of energy versus volume ($EV$) relations for
various crystal structures. For Al, $EV$ curves were computed here for the
face centered cubic (fcc), hexagonal closed packed (hcp), body centered
cubic (bcc), simple cubic (sc), diamond, and the \l10 structures.  The
$L1_0$ structure of Al is a defected fcc lattice with a vacancy in the
corner of each cubic unit cell.  In the case of titanium, $EV$ curves were
generated in Ref.~\onlinecite{Mehl02a} for the hcp, fcc, bcc, sc, and the
omega (C32) structures.  For the intermetallic compound TiAl, $EV$ relations
were obtained in this work for three structures: \l10 (CuAu prototype), B2
(CsCl prototype), and B1 (NaCl prototype). Each $EV$ curve typically
consists of total energies for about $20$-$30$ different volumes around the
equilibrium volume. The $c/a$ ratio of the \l10 structure has been optimized
at each volume.  As the \abinit and EAM energies are at different scales,
all \abinit energies for a given element or compound of a given
stoichiometry were shifted so that to match the experimental cohesive energy
of the equilibrium ground state structure. This procedure is followed merely
to facilitate the comparison of structural energies calculated by different
methods and does not introduce any new approximation. The equilibrium energy
of the D0$_{19}$-Ti$_3$Al compound has also been calculated here to check
the methodology and provide useful reference information. The energy of this
phase was minimized with respect to its volume while keeping the $c/a$ ratio
fixed at the value found in Ref.~\onlinecite{WW98}.

The $EV$ curves were calculated using the full potential LAPW
method\cite{anderson75,WK85,Dsingh} within the Hohenberg-Kohn-Sham
formulation of the density functional theory.\cite{HK64,KS65,PY89} The
calculations were carried out in a spin-restricted mode and the
exchange-correlation effects were treated at the level of the GGA. The $EV$
calculations for Ti were carried out using an LAPW code available at the
Naval Research Laboratory and were reported in Ref.~\onlinecite{Mehl02a}. The
calculations for Al, TiAl and Ti$_3$Al were performed in this work and
employed the {WIEN2K} package.\cite{wien2k} The
Perdew-Wang\cite{PWA91,PWB91,PWC91} (PW91) exchange-correlation functional
was used for Ti calculations,\cite{Mehl02a} while its simplified and more
efficient version referred to in literature as PBE\cite{PBE} was used for Al
and TiAl.  The muffin-tin radii of Al and Ti were chosen to be 2 a.u.  For
each crystal structure, systematic \vk point and basis set convergence tests
were carried out at a fixed volume near equilibrium volume. The same set of
parameters was subsequently employed for different volumes as well as for
the Bain path calculations. The accuracy of the LAPW total energies
calculated in the present work was estimated to be better than 0.5 mRy/atom.

We now describe some relevant details of the potential fitting procedure.
In the embedded atom formalism,\cite{DB83_0,DB84} the total energy of a
system is expressed as
\begin{equation}
    E_{tot} = \frac{1}{2} \sum_{i,j}\Phi_{ij} (r_{ij}) + \sum_{i}
    F_i(\bar{\rho_i}). 
                  \label{eq:EAM}
 \end{equation}
Here, $\Phi_{ij}$ is the pair-interaction energy between atom $i$ and $j$ at
positions $\vec{r}_i$ and $\vec{r}_j$, and $F_i$ is the embedding energy of
atom $i$.  The $\bar{\rho}_i$ in Eq.~(\ref{eq:EAM}) is the host electron
density at site $i$ induced by all other atoms in the system. The latter
is given by
\begin{equation}
 \bar{\rho_i} = \sum_{j\neq i} \rho_j(r_{ij}).
\end{equation}
For a binary system A-B, the total energy given by Eq.~(\ref{eq:EAM}) is
invariant with respect to the following set of
transformations:\cite{Johnson89,MMP02}
\begin{eqnarray}
&&  \rho_A(r)  \rightarrow  s_A \rho_A(r), \\
&&  \rho_B(r)  \rightarrow  s_B \rho_B(r),      \\
&&  F_A (\bar{\rho}) \rightarrow  F_A(\rho_A(r)/s_A),      \\
&&  F_B (\bar{\rho}) \rightarrow  F_B(\rho_B(r)/s_B),      \\
&&  F_A (\bar{\rho}) \rightarrow  F_A(\bar{\rho}) + g_A\bar{\rho},      \\
&&  F_B (\bar{\rho}) \rightarrow  F_B(\bar{\rho}) + g_B\bar{\rho},      \\
&& \Phi_{AA}(r) \rightarrow  \Phi_{AA}(r) - 2 g_A \rho_A(r),      \\
&& \Phi_{BB}(r) \rightarrow  \Phi_{BB}(r) - 2 g_B \rho_B(r),      \\
&& \Phi_{AB}(r) \rightarrow  \Phi_{AB}(r) -2 g_A \rho_A(r)- 2 g_B \rho_B(r),      
\end{eqnarray}
where $A$ and $B$ refer to the type of element (Al or Ti) and $s_A, s_B,
g_A, \mbox{\,and\,} g_B$ are arbitrary constants. For any particular
compound, the potential can be cast into the so-called effective pair
format\cite{Johnson89} by choosing $g_A=-F'_A(\bar{\rho_A}^0)$ and
$g_B=-F'_B(\bar{\rho_B}^0)$, where $\bar{\rho_A}^0$ and $\bar{\rho_B}^0$ are
the equilibrium electron densities on atoms in the compound. The effective
pair format provides a convenient way of comparing different potentials for
the same compound.

An EAM potential for a binary system can be constructed by two different
procedures. One is to optimize all potential functions simultaneously
during a single fit, as it was done for example for the Ni-Al
system.\cite{MMP02} This scheme offers the advantage of having many
parameters available for fitting to properties of a selected alloy or
compound.  However, it suffers from the drawback that the quality of
potentials for pure elements (Al and Ti in our case) is often poor. An
alternative and more common approach is to separately develop accurate EAM
potentials for pure elements and use them to fit the cross potential for
alloys (compounds). We resort to the latter procedure for the Ti-Al system.
The parametrization of potential functions employed in the fitting procedure
is discussed below.

\subsection{\label{sec:AlEAM}EAM potential for Al}

For the EAM potential for Al, we chose the electron density function in the
form
\begin{eqnarray}
 \rho(r) = \psi \biggl (\frac{r-r_c}{h} \biggr ) \Bigl \{ A_0(r-r_0)^{y} 
    e^{ -\gamma(r-r_0)} \nonumber  \\
  \times ~\Bigl (1+B_0 \,  e^{ -\gamma(r-r_0)} \Bigr ) + C_0
  \Bigr \},
 \end{eqnarray}
Here, $A_0,B_0,C_0,r_0, r_c,h,y \mbox{ and } \gamma $ are the fitting
parameters and $\psi(x)$ is a cutoff function defined as
\begin{eqnarray}
\psi(x) & = &  0   ~~{for~~} x\ge 0  \nonumber \\
        & = &   \frac{x^4}{(1+x^4)} ~~{for~~} x < 0,  
                \label{eq:cutoff}
\end{eqnarray}
where $r_c$ is the cutoff distance.  The electron density in equilibrium fcc
Al is normalized to unity, i.e. $\bar{\rho} = \sum_j N_j \rho_j = 1$ , where
$j$ runs over coordination shells and $N_j$ is the number of atoms on the
$j$-th coordination shell. This constraint fixes one parameter in the
above set of parameters. The pair interaction function is parametrized in
the form
\begin{eqnarray}
  \Phi(r) = \Biggl [ \frac{V_0}{(b_2-b_1)} \biggl (\frac{b_2}{z^{b_1}} - \frac{b_1}{z^{b_2}} \biggr )
   + \delta \Biggl ]  \psi \biggl (\frac{r-r_c}{h} \biggr ),  
         \label{eq:pairAl}
\end{eqnarray}
where, $z=r/r^{\prime}$, and $b_1$, $b_2$, $\delta$, $V_0$, and $r^{\prime}$
are fitting parameters.  Thus, altogether we have $12$ fitting parameters
for functions $\rho(r)$ and $\Phi(r)$.

The embedding energy is obtained by equating the energy of fcc Al
[Eq.~(\ref{eq:EAM})] to the universal equation of state (EOS). By studying a
broad range of materials, Rose \etal\cite{Rose} proposed the ``universal''
EOS in the form
\begin{equation}
 E(r) = -E_{0} \bigl [ 1 + \alpha (r/r_e -1) \bigr ] e^{-\alpha(r/r_e-1)},
       \label{eq:Rose}
\end{equation}
where $\alpha = \sqrt{9\Omega_0 B/E_0}$, and $r$, $E_0$, $\Omega_0$, $B$,
$r_e$ are the nearest-neighbor distance, cohesive energy, equilibrium atomic
volume, the bulk modulus, and the equilibrium-nearest neighbor
distance, respectively. It is generally found that the EAM potentials which
exactly follow Rose's EOS [Eq.~(\ref{eq:Rose})] underestimate energies at
high pressures.\cite{MMP02}  We have therefore modified Eq.~(\ref{eq:Rose})
to allow for a more accurate fit to energies at high pressures. The modified
equation has the form
\begin{equation}
 E(r) = -E_{0} \Biggl [ 1 + \alpha x + \beta \alpha^3 x^3
\frac{2 x + 3}{(x+1)^2}\Biggr ] e^{-\alpha x},
       \label{eq:geos}
\end{equation}
with $x = (r/r_e -1)$.  The parameter $\beta$ in this equation is related to
the pressure derivative of the bulk modulus at equilibrium as $B_0^\prime =
\frac{2}{3}\alpha + 6\alpha\beta + 2$.  This modification does not alter the
exact fitting of the potential to $E_0, r_e$ and $B$ but provides a way to
achieve an accurate fit to the experimental pressure-volume relation by
adjusting the value of $\beta$.

\subsection{\label{sec:TiEAM}EAM potential for Ti}

Titanium has the hcp structure at $T=0$ ($\alpha$-Ti phase).  For hcp
metals, an EAM potential can only be fitted to elastic constants $C_{ij}$ if
the relation $(3C_{12}-C_{11})/2 > (C_{13}-C_{44})$ is satisfied.\cite{PS92}
Fortunately, this relation holds for $\alpha$-Ti.

In the present EAM potential for Ti, the electron density function is
described by
\begin{eqnarray}
\rho(r) = \Biggl [A e^{-\alpha_1(r-r_{0})^2} + e^{-\alpha_2(r-r_0^{\prime})}
  \Biggr ] \psi \biggl (\frac{r-r_c}{h} \biggr ), \label{eq:Tieden}
\end{eqnarray}
where the cutoff function $\psi(x)$ is given by Eq.~(\ref{eq:cutoff}) with
the fitting parameters $A,\alpha_1, \alpha_2, r_0, r_0^{\prime}, r_c$, and
$h$.  One of the parameters is fixed by the normalization condition
$\bar{\rho} = 1$ at equilibrium.

The pair interaction function is represented by 
\begin{eqnarray}
\Phi(r_{ij}) = \psi \biggl (\frac{r-r_c}{h} \biggr )
\biggl  \{
 V_0 \,e^{-\beta_1 r_1} +
 V_0^{\prime} \Bigl [ e^{-2  \beta_2(r-r_1^{\prime})} - 2 \, e^{\beta_2(r-r_1^{\prime})} \Bigr ] + \delta
\biggr \},
         \label{eq:Tipair}
\end{eqnarray}
where $V_0,V_0^{\prime},\beta_1,\beta_2,r_1, r_1^{\prime}, \mbox{\,and\,}
\delta$ are fitting parameters.  The embedding energy function is expressed
as a polynomial:
\begin{eqnarray}
F(\rhm) &=& F_{0} + \frac{1}{2}F_2 (\rhm-1)^2 + q_0 (\rhm -1)^3 \nonumber \\ 
 & & + ~\sum_{i=1}^{3} B_i (\rhm-1)^{i+3}.
         \label{eq:Tiembed}
\end{eqnarray}
Here $F_0$ and $F_2$ are the embedding energy and its second derivative at
equilibrium, respectively.  These can be expressed in terms of the
experimental values of $E_0$, $B$, and $\Omega_0$:
$$ 
F_0 = E_0 - \frac{1}{2} \sum_j N_j \Phi_j 
$$
and
$$ 
\frac{1}{2} \sum_j N_j \Phi^{\prime\prime}_j R^2_j + F_2 \Bigl (\sum_j N_j
\rho_j R_j\Bigr )^2 = 9B\Omega_0,
$$
where $j$ runs over coordination shells, $N_j$ is the number of atoms on the
$j$-{th} coordination shell of radius $R_j$, while $\Phi_j$ and
$\Phi_j^{\prime\prime}$ are the pair interaction energy and its second
derivative evaluated at $R_j$.  The coefficients $q_0$ and $B_i ~(i\le 3)$
in Eq.~(\ref{eq:Tiembed}) are fitting parameters. The parameter $q_0$ was
adjusted to ensure that the embedding energy vanishes when the electron
density goes to zero, that is, $F(0) = 0$.  This requirement leads
to the following expression for $q_0$:
\begin{eqnarray}
q_0 = F_{0} + \frac{F_2}{2} + B_1 - B_2 + B_3.
\end{eqnarray}
During the optimization of fitting parameters, the energy of the hcp
structure was required to approximately follow the Rose's EOS
[Eq.~(\ref{eq:Rose})] in the neighborhood of equilibrium.  This was achieved
by adding to the objective function the mean-squared deviation of the energy
from Eq.~(\ref{eq:Rose}) at several points near the equilibrium.

\subsection{\label{sec:TiAlEAM}The cross potential Ti-Al and the fitting
procedure}

Once the EAM potentials for Al and Ti are obtained, the cross potential
representing the interactions between Ti and Al atoms was constructed by
employing the parametrization given by Eq.~(\ref{eq:pairAl}).  The
transformation coefficients $s_{Al}, g_{Ti},$ and $g_{Al}$ [see
Eqs.~(3)-(11)] were used as additional adjustable parameters.

\begin{figure}
\includegraphics[scale=0.9]{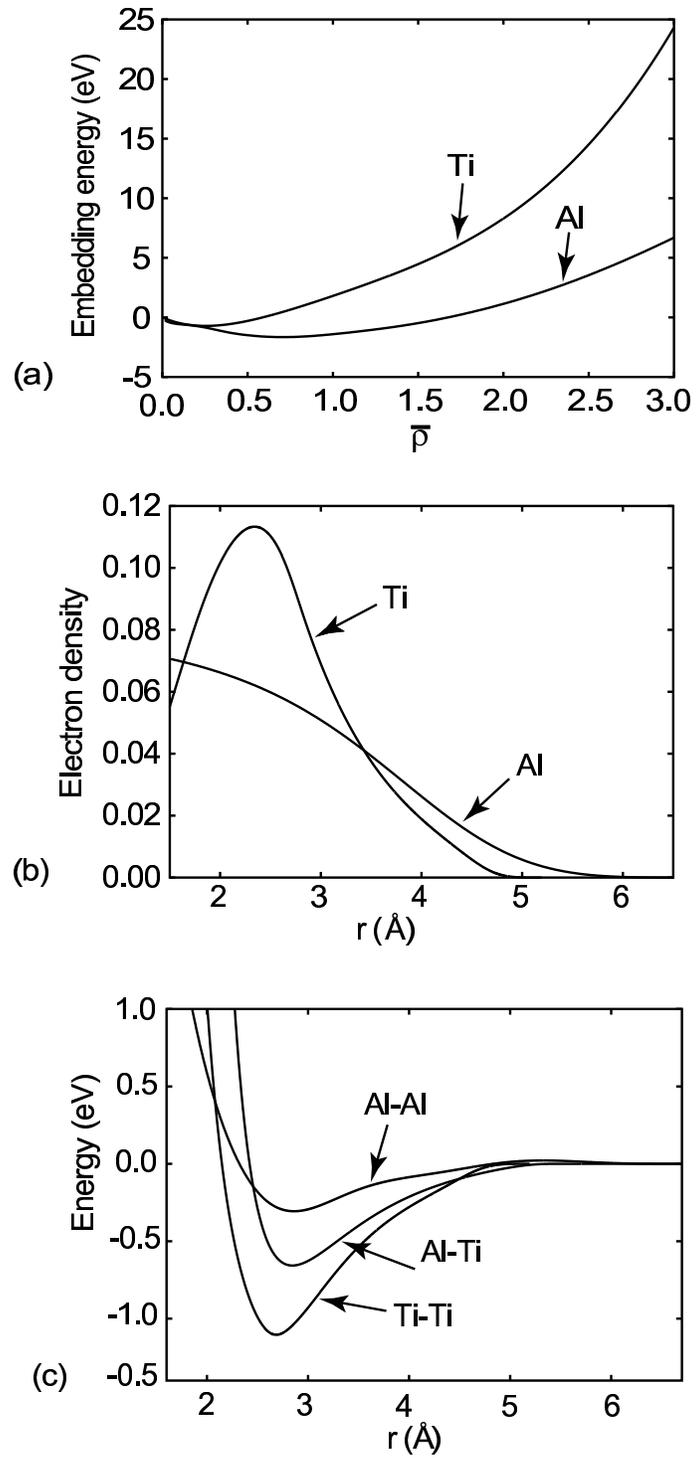}
\caption{\label{func:fig} The embedding energy (a), electron density (in arbitrary units) (b),
and the pair interaction function (c) for the Ti-Al system in the effective
pair format.}
\end{figure}

It should be mentioned that the specific analytical forms of the potential
functions adopted in this work were found by trying a number of different
forms and selecting those which provided a better accuracy of fitting with
less parameters.  The optimized values of the fitting parameters are listed
in Table~\ref{tab:param}.  The potential functions are plotted in
Fig.~\ref{func:fig} in the effective pair format\cite{MMP02,Johnson89} with
respect to \gtial.  These functions are available in the tabulated form on
Internet\cite{NRL} or from the authors upon request. The cutoff radii of
atomic interactions in Al, Ti, and TiAl are 6.72, 5.19, and 5.77 \AA,
respectively. The fitting procedure involves the total of $39$ independent
fitting parameters.

\section{Tests and Applications of the potentials}

\subsection{Aluminum}

The accuracy of the fitted EAM potential for Al can be adjudged from
Table~\ref{aleq:tab}, wherein the basic lattice properties, elastic
constants, vacancy formation and migration energies, surface energies, and
the stacking fault energy predicted by the potential are compared with
their experimental values.  We have also included the results obtained with
our previous EAM potential for Al,\cite{MFMP99} which we hereafter refer to
as MFMP.  The results obtained with the new EAM potential are in good
agreement with their experimental counterparts. The calculated vacancy
formation energy $E^f_v$ and migration energy $E^m_v$, which are important
for studying point defect diffusion, are well reproduced.
The calculated vacancy formation volume $\Omega_v^f$ compares well with the
one obtained with the MFMP potential as well as with experimental
data.\cite{Emrick69} 
The predicted intrinsic stacking fault $(\gamma_{SF})$ energy is on the lower
end of the range of experimental values  while the 
{\it ab initio}\cite{Kaxiras97, Kaxiras00} calculatations yeilds a higher value of 
 166 \mjm2 .
The  calculated symmetrical twin boundary $(\gamma_T)$ energy 
is in good agreement with its experimental counterpart. 
The unstable stacking 
fault energy is  underestimated with respect to  the
{\it ab initio}\cite{Kaxiras97, Kaxiras00} value of 220 \mjm2.
The surface energies are
underestimated in comparison with experiment, which is a general
characteristic of EAM potentials.

\begin{figure}
\includegraphics{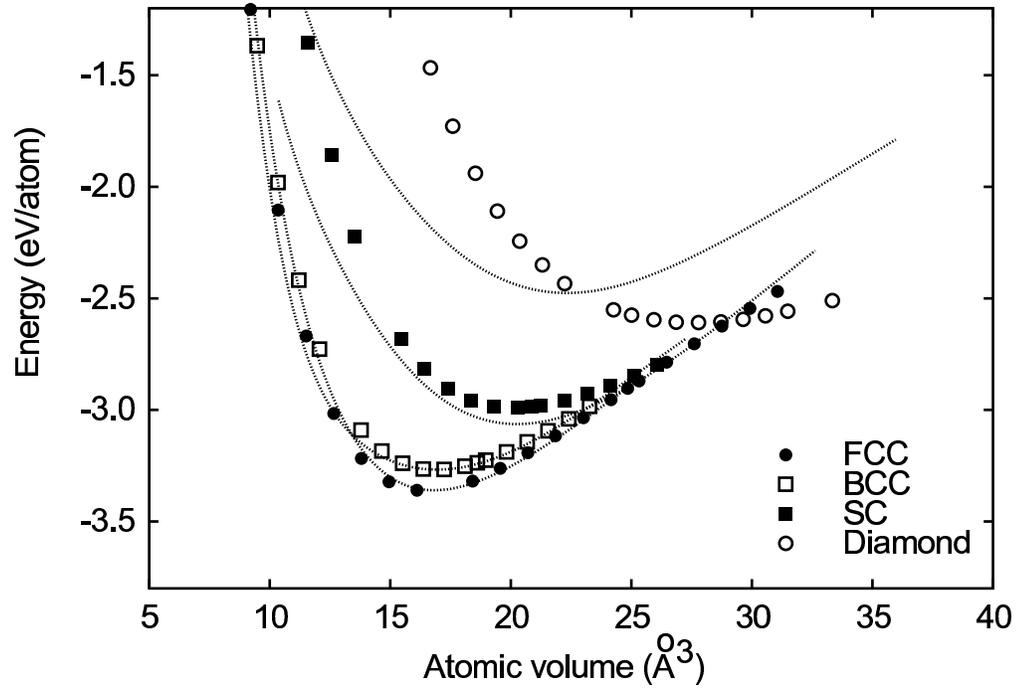}
\caption{ Energy-volume relations for different crystalline structures of Al calculated with  the EAM potential (lines) and by 
the LAPW method (points).}
\label{aleos:fig}
\end{figure}

The structural energy differences for Al, given in Table~\ref{edfal:tab},
reproduce \abinit energies reasonably accurately. The EAM potential predicts
the $c/a$ ratio of the hcp structure to be 1.63 in good agreement with the
optimized value of 1.645 obtained by the LAPW calculations.  In agreement
with earlier findings,\cite{MFMP99} bcc Al is mechanically unstable ($C_{11}
< C_{12}$) and transforms to the fcc structure upon $c/a$ relaxation. The sc
structure is also mechanically unstable with $C_{44} < 0$. A comparison of
the EOS's of various crystalline structures of Al calculated with the EAM
potential and by the LAPW method is presented in Fig.~\ref{aleos:fig}. The
EAM curves are seen to agree with the LAPW results fairly well. The
agreement is particularly good for the bcc and fcc structures over a large
range of volumes, but tends to worsen for more open structures.

\begin{figure}
\includegraphics[scale=0.8]{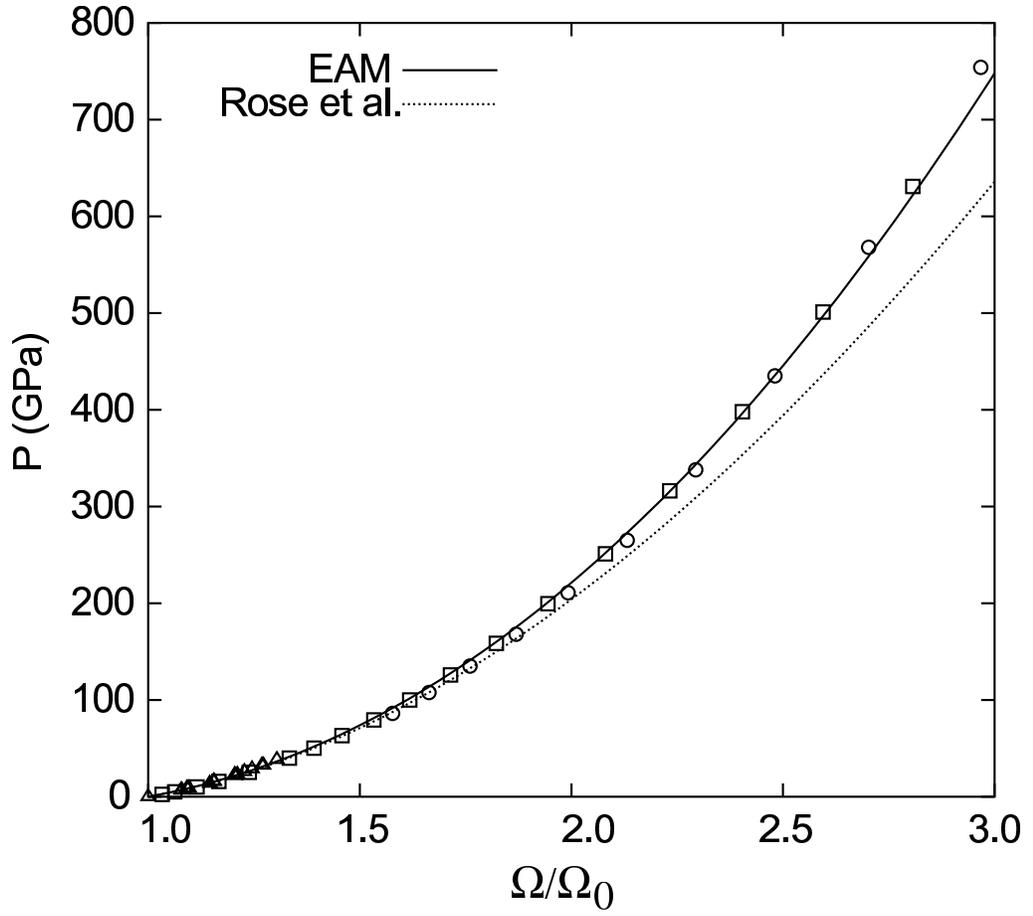}
\caption{ 
The pressure-volume relation for aluminum at $T=0$ calculated with the present
EAM potential (solid line), predicted by the universal equation of state\cite{Rose} (dotted line),
and measured experimentally (squares: Ref.~\onlinecite{WCZ00};  circles:
Ref.~\onlinecite{AlEOS2}; triangles: Ref.~\onlinecite{AlEOS3}).}
\label{alPV:fig} 
\end{figure}

As was mentioned in Section II B, the Al potential was fit to the
experimental $P(V)$ relation by adjusting the parameter $\beta$ in the
generalized EOS, Eq.~(\ref{eq:geos}). The optimized value of $\beta =
0.00489$ provides an excellent agreement with experimental data up to
pressures of about 700 GPa, as illustrated in Fig.~\ref{alPV:fig}. In
contrast, the standard Rose's EOS ($\beta=0$) underestimates the pressures
under strong compressions. Note that both equations share the same values of
$E_0$, $B$ and $a_0$. The excellent fit to high pressure data makes the
potential useful for simulating shock waves, sputter deposition, and other
processes involving a close approach of atoms.

Thermal expansion of Al was studied within the temperature range of
$5$-$1000$ K. The calculated thermal expansion factors at selected
temperatures are given in Table~\ref{teal:tab}.  They were obtained using
a 864-atom supercell by two different methods. In the first method, the free
energy of the crystal was minimized as a function of volume in the
quasiharmonic approximation.\cite{Foiles94} This method includes
quantum-mechanical effects such as zero-point vibrations and should yield
more accurate values of the thermal expansion factor at low
temperatures. However, it may not be very accurate at high temperature where
the anharmonic effects become significant.  The second type of calculation
was carried out by the Metropolis Monte Carlo method.\cite{Foiles94,FA89}
This method is based on classical mechanics and fully incorporates anharmonic
effects. It is therefore more adequate for thermal expansion calculations
at high temperatures.  As can be seen from Table~\ref{teal:tab}, the MC
results are quite close to experimental data at high temperatures.

Overall, the EAM potential developed here provides a good description of a
wide range of Al properties. Despite the existence of other high-quality EAM
potentials for Al in the literature,\cite{EA94,HVF99,MFMP99} we chose not to
re-use one of them but rather generate a new potential so that to have all
potential functions for the Ti-Al system created by the same methodology. We
also used this work as an opportunity to address some weak points of previous
potentials. For example, even though the MFMP potential\cite{MFMP99}
demonstrates a better agreement with experiment for some of the properties
listed in Table~\ref{aleq:tab}, the present potential describes the thermal
expansion and high-pressure behavior of Al more accurately. The present
potential is also based on a larger set of \abinit data and should be better
transferable to configurations away from equilibrium. It should also be
mentioned that the use of smooth analytical functions in this work makes the
potential more robust in comparison with the cubic-spline parameterization
applied in Ref.~\onlinecite{MFMP99}.

\subsection{Titanium}

Equilibrium lattice properties, vacancy characteristics, as well as stacking
fault and surface energies in Ti computed using the present EAM potential 
are compared with experimental data in Table~\ref{tieq:tab}.  We
have also included the results obtained with the EAM potential developed by
Fernandez \etal,\cite{FMP95} which will be referred to as FMP. The latter is
an improved version of the potential proposed in
Ref.~\onlinecite{PS92}. Although our potential was not fitted to the $c/a$
ratio exactly, the predicted value of 1.585 is in a good agreement with the
experimental value of 1.588.  The elastic constants are also reproduced
reasonably well.

The vacancy formation energy $E_v^f$ was fitted to the target value of
$1.85$ eV. The experimental value of \evf reported by
Shestopal\cite{Shestopal66} is $1.55$ eV, more recent positron annihilations
measurements\cite{HSK84} give $ E_v^f=1.27$ eV, while the \abinit linearized
muffin-tin orbital method (LMTO) method\cite{BWP99} yields a much higher
value of $2.14$ eV. We therefore opted to fit to an intermediate value of
$1.85$ eV, which after the relaxation decreased to $1.83$ eV.  The vacancy
migration energy $E^m_v$ was calculated using the nudged elastic band
method.\cite{NEB} In the hcp lattice with a nonideal $c/a$ ratio, the basal
and non-basal vacancy jumps are not identical. The calculated values of \evm
for the basal and non-basal jumps are $0.80$ eV and $0.83$ eV,
respectively. The FMP potential gives smaller values of $0.51$  eV and $0.48$
eV, respectively. While the experimental value of the vacancy formation
energy is rather uncertain, the experimental activation energy $Q$ of
self-diffusion in \alphati, which is the sum of \evf and \evm, has been
measured fairly accurately.\cite{KHFM97} For self-diffusion perpendicular to
the $c$ axis, the experimental value is $Q=3.14$ eV. The present EAM
potential predicts $Q=2.62$ eV, while the FMP potential gives a lower value
of $2.02$ eV.

There are three stacking faults on the basal plane in \alphati, which are
deviations from the normal stacking sequence ${ABABAB}$ of closed packed
planes in the hcp structure.\cite{HB,HL82}  The intrinsic fault $I_1$ is
formed by a removal of one hexagonal layer followed by a
$\frac{1}{3}<10\bar{1}0>$ translation of all atoms above this fault. The
resultant stacking sequence is
$$
I_1: ~~\,\,ABAB{\vert}CBCBC , 
$$  
where the vertical bar indicates the position of the fault.  The intrinsic
stacking fault $I_2$ is created by a $\frac{1}{3}<10\bar{1}0>$ slip:
$$
I_2: ~~\,\,ABAB\vert CACAC . 
$$
The extrinsic stacking  fault $E$ result from the insertion of an extra
hexagonal plane into the normal stacking sequence:
$$
I_E:~ ~\,\,ABAB{\vert}C{\vert}ABAB .
$$ 
The calculated relaxed stacking fault energies (Table~\ref{tieq:tab})
compare well with those obtained with the FMP potential. The experimental
value of the $I_2$ fault energy is about $300$ \mjm2 and is considered to be
a rough estimate. Both EAM potentials underestimate this experimental
value. All our efforts to obtain a higher $\gamma_{I_2}$ value during the
fitting of the present potential did not have much success. In fact, any
attempt to raise $\gamma_{I_2}$ above $66$ \mjm2 resulted in a deterioration
of other properties, which gave us an indication that higher stacking fault
energies may be beyond the capabilities of the EAM.  Note, however, that the
EAM-predicted stacking fault energies follow the expected relation:\cite{HB}
$\gamma_E \approx \frac{3}{2} \gamma_{I_2} \approx 3 \gamma_{I_1} $.  The
limited success in fitting to higher stacking fault energies is likely to be
due to the directional component of bonding in Ti owing to d electrons. The
covalent nature of bonding cannot be described by the central-force-based
EAM model. More rigorous parameter-based methods such as the modified
EAM,\cite{BJ94,MEAM} bond order potentials\cite{bop1,bop2,bopti} or the
tight-binding method,\cite{MP96} which include angular-dependent
interactions, may give higher stacking fault energies.

The predicted value of the $(0001)$ surface energy, $1725$ \mjm2, slightly
underestimates the experimental value. This is again consistent with the
general trend of the EAM to underestimate surface energies.  The FMP
potential yields an even smaller value of $1439$ \mjm2, whereas the \abinit
surface energies, $2100$ \mjm2 (Ref.~\onlinecite{BBMMN88}) and 1920 \mjm2
(Ref.~\onlinecite{TM77}), overestimate the experimental value.

\begin{figure*}
\includegraphics[scale=0.9]{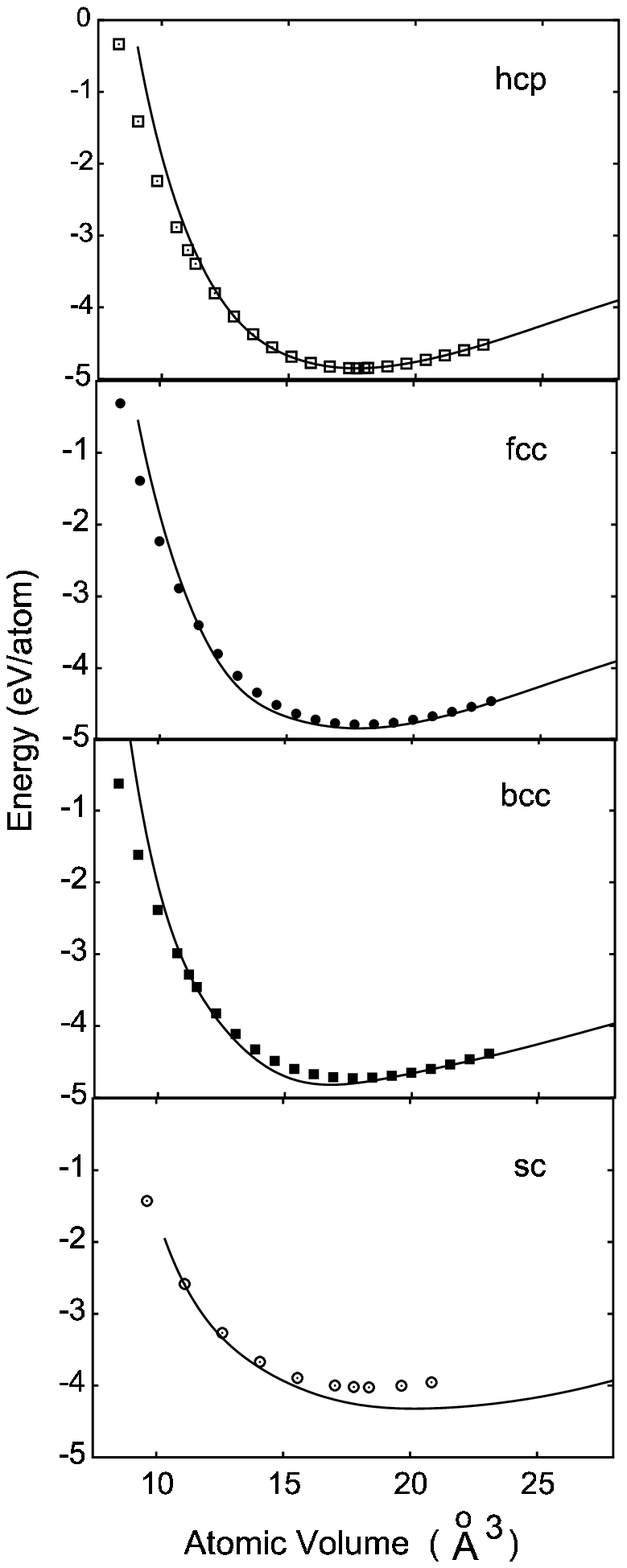} 
\caption{\label{tieos:fig} Comparison of the LAPW (points) and  EAM (solid
lines) energy-volume relations for different structures of Ti. The LAPW
energies were calculated in Ref.~\onlinecite{Mehl02a}.} 

\end{figure*}

The LAPW and the EAM energies of various crystal structures of Ti relative
to the hcp structure are reported in Table~\ref{edfti:tab}. We note that the
LAPW calculations predict the omega structure to be the ground state, with
the hcp energy being 0.06 eV/atom higher. We have, therefore, excluded the
omega structure from the fitting procedure. Overall, both EAM potentials
yield similar energy differences between the structures, with the present
potential performing somewhat better. Both potentials predict the hcp
structure to be more stable than the omega structure in agreement with
experiment. In Fig.~\ref{tieos:fig}, the EOS's of the hcp, fcc, bcc, and the
sc structures of Ti calculated with the present EAM potential are compared
with the LAPW results. The agreement between the two calculation methods is
good for the close-packed structures but becomes poorer for the low
coordinated sc structure. In the latter case, however, the present EAM
potential is closer to the LAPW data than the FMP potential.

The linear thermal expansion of Ti was calculated within the quasiharmonic
approximation and by the MC method using a supercell with 800 atoms.  The
$c/a$ ratio was kept fixed at its equilibrium $T=0$ value during the
calculations.  The obtained values of the thermal expansion factor for
selected temperatures are reported in Table~\ref{teti:tab}. The agreement
with experimental data\cite{TKTD75} for polycrystalline Ti is reasonable.
The FMP potential gives a poorer agreement with experiment.  For example, at
293 K the FMP potential gives the quasiharmonic linear thermal expansion of
1.35\% while the experimental value is 0.15\%.

\subsection{Intermetallic compound $\gamma$-TiAl}

The physical properties of \gtial obtained with the present EAM potential
are summarized in Table~\ref{eqtial:tab}.  The lattice constant and the
cohesive energy are reproduced accurately.  The $c/a$ ratio is correctly
predicted to be larger than unity and is in good agreement with the
experimental value. The elastic constants are in a reasonable agreement with
experiment, the root-mean-squared deviation for elastic constants being
about 22\%. We note that the negative signs of the two Cauchy pressures,
($C_{12}-C_{66}$) and ($C_{13}-C_{44}$), are not reproduced by the present
potential, nor are they reproduced by previous EAM-type potentials. The
negative Cauchy pressures in TiAl are caused by the directional component of
bonding and cannot be described by the EAM. Table \ref{eqtial:tab} also
includes the numbers calculated with the Farkas
potential\cite{Farkas94,Farkas-elast} as well as with the P$_2$ potential
constructed by Simmons \etal \cite{SRD97} The respective root-mean-squared
deviations of the elastic constants from experimental data are 45\% and
28\%.

The planar defect energies in \gtial are summarized in
Table~\ref{sftial:tab}. They were calculated using supercells with an
effective cubic lattice with the lattice parameter
$a=\left(a_0^2c_0\right)^{1/3}$ and without volume relaxation. Since the cohesive energy of the
equilibrium tetragonal lattice was used as a reference in the calculations,
the resultant fault energies can be slightly overestimated. For the
superlattice intrinsic stacking fault (SISF), the EAM value of $173$ \mjm2
is slightly higher than the experimental value of $140$ \mjm2.  There have
been a number of \abinit calculations of the SISF energy with results
scattered over the range of $90$-$172$ \mjm2. For the complex stacking fault
(CSF) energy we obtain the value of $299$ \mjm2 well bracketed between \abinit
results. Experimental data for the CSF energy are not available.  The
calculated value of the antiphase boundary (APB) energy, $266$ \mjm2, is
also in a good agreement with the experimental value of $250$ \mjm2. The
\abinit APB energies are scattered over the wide range $510$-$670$
\mjm2. The hierarchy of planar fault energies in \gtial was investigated by
Wiezorek and Humphreys.\cite{WH95} According to their preliminary
computational results, this hierarchy in Ti-$54$at\%Al is $\gamma_{CSF} >
\gamma_{APB} > \gamma_{SISF}$. The present EAM potential predicts the same
ordering. On the other hand, \abinit calculations with the linearized
Korringa-Kohn-Rostoker (LKKR) method\cite{WMR92} and the full potential LAPW
(FLAPW) method\cite{FY90} give the $\gamma_{APB} >\gamma_{CSF} >
\gamma_{SISF}$ ordering. More recent calculations by Ehmann and F\"ahnle by
the LAPW method including local atomic relaxations are consistent with the
latter ordering of the stacking fault energies. The discrepancy between the
experimental data and EAM calculations, on one hand, and \abinit
calculations, on the other hand, originates primarily from the high APB
energy delivered consistently by \abinit methods. The low APB energy observed
experimentally may reflect the local disorder near the APB taking place due
to the off-stoichiometry and/or temperature effects. The similarly low APB
energy predicted by the present EAM potential can lead to a good agreement
between atomistic simulations and experiment.

Simmons \etal\cite{SRD97} succeeded in generating a set of EAM potentials
for \gtial fit to high APB energies comparable to \abinit values, but their
potentials give $c/a<1$ in contradiction to experimental data. When
generating our potential we could also achieve higher $\gamma_{APB}$ values
at the expense of $c/a<1$, but could never increase $\gamma_{APB}$ above
$266$ \mjm2 while keeping $c/a>1$ and maintaining a good quality of fit to
other properties. We believe that the underestimation of the APB energy is
another intrinsic limitation of the central-force EAM as applied to
\gtial. Farkas\cite{Farkas94} constructed a potential that gives
$\gamma_{APB} > \gamma_{CSF}$ while $c/a>1$. However, some of the elastic
constants predicted by that potential are in a poor agreement with
experimental data, especially $C_{13}$ and $C_{33}$ (cf.\
Table~\ref{eqtial:tab}). The potential also gives a discontinuous
temperature dependence of the quasiharmonic thermal expansion, with
unrealistically large values at high temperatures.

\begin{figure}
\includegraphics{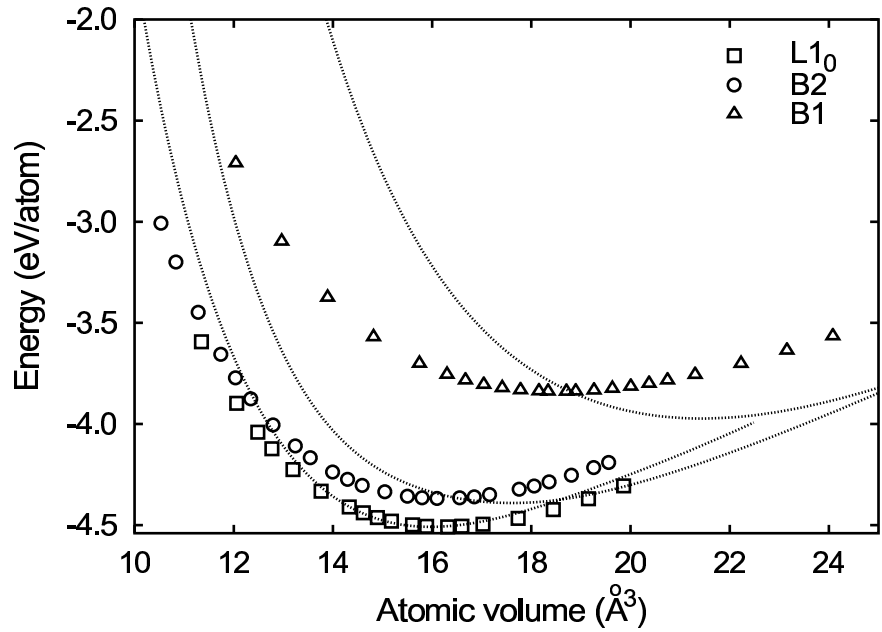}% 
\caption{ Energy-volume relations for the $L1_0$, B2 and B1 structures of
TiAl.  The LAPW results are marked by points, the lines represent EAM
calculations.}
\label{tialeos:fig} 
\end{figure}

      The $EV$ curves computed with the present EAM potential and by the
LAPW method are presented in Fig.~\ref{tialeos:fig}. The agreement between
the two calculation methods is good for the L1$_0$ structure.  
The difference between the two curves for the B2 and B1 structures is presumably due to
the limited accuracy of the EAM method to describe the open structures.
The formation energies for
different structures of TiAl (relative to fcc Al and hcp Ti) calculated by
the LAPW method and with the present EAM potential are presented in
Table~\ref{eform:tab}. For comparison, experimental and \abinit results
available in the literature have also been included in the Table. The
formation energies obtained with the present EAM potential are in good
agreement with the corresponding experimental and \abinit energies. The \l10
structure is correctly produced to be the ground state. We note the $B2$ and
$B32$ structures are unstable with respect to the $c/a$ optimization. In
particular, the $B2$ structure transforms to the equilibrium \l10 phase
upon $c/a$ relaxation.

For \ti3al, the EAM potential correctly predicts the DO$_{19}$ structure to
be the equilibrium ground state of \ti3al (Table~\ref{eform:tab}). The
lattice constants, cohesive energy, and the elastic constants of
D0$_{19}$-\ti3al are given in Table~\ref{ti3al:tab}.  We emphasize that none
of these properties were included in the potential fit. The observed
agreement with experimental data demonstrates a good transferability of our
potential.

The EAM potential was also applied to investigate the \al3ti
compound. Experimentally, the equilibrium structure of \al3ti is
DO$_{22}$. The present EAM potential predicts the L1$_{2}$
structure to be $0.01$ eV lower in energy than the DO$_{22}$ structure,
suggesting that the potential may not suitable for simulating the \al3ti
compound.

Thermal expansion factors of \gtial calculated
within the quasiharmonic approximation and by the Monte Carlo method are
presented in Table~\ref{te:tab}.  The calculated values are in agreement
with those estimated from Fig.~4 in Ref.~\onlinecite{ZH01}.

\begin{figure}
\includegraphics{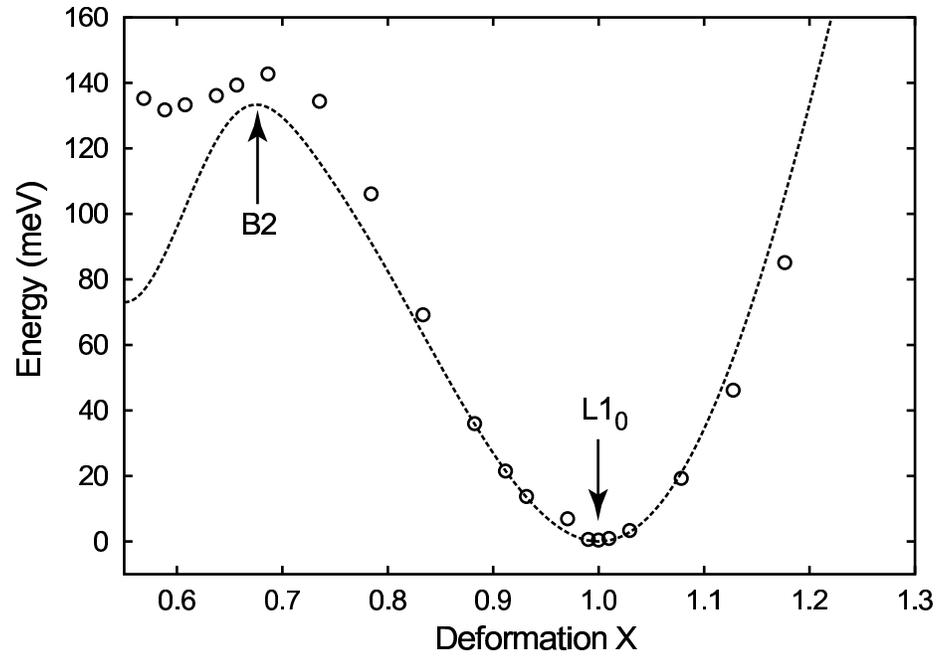}% 
\caption{
Energy per atom as a function of the deformation parameter X (see text for details) along the
volume conserving  tetragonal deformation
path  (Bain path) in TiAl. 
The energy is given relative to the equilibrium $L1_0$ structure.
The dotted line is predicted by the present EAM potential and the symbols represents the LAPW results. }
\label{bain:fig}
\end{figure}

The energy along the Bain path between the tetragonal \gtial structure and
the B2 structure was calculated by the EAM and LAPW methods.  Starting from
the equilibrium tetragonal \gtial structure, the $c/a$ ratio was varied by
keeping the volume constant. The energy change during the transformation is
plotted in Fig.~\ref{bain:fig} as a function of the deformation parameter $X$
defined by $c/a=X(c_0/a_0)$.  The EAM energies are observed to closely
follow the LAPW energies along the path. This agreement confirms a good
transferability of the present EAM potential.

Point defect properties play an important role in the atomic disorder and
diffusion in \gtial. The TiAl lattice supports two types of vacancy (V$_{\rm
Ti}$ and V$_{\rm Al}$) and two types of antisite defects (Ti atom on the Al
sublattice, Ti$_{\rm Al}$, and Al atom on the Ti sublattice, Al$_{\rm
Ti}$).\cite{MH00} The so-called ``raw" formation energies and
entropies\cite{MH00} of the defect formation have been calculated with the
present EAM potential using the molecular statics method for the energies
and the quasiharmonic approximation for the entropies.

When analyzing point defects in ordered compounds it is more convenient to
deal with hypothetical composition-conserving defect complexes rather than
individual defects.\cite{MH00,Hagen98,Pasha2000,Fahnle99} It should be
emphasized that the defects are grouped into complexes conceptually and not
physically. The complexes are assumed to be totally dissociated and
interactions between their constituents are neglected. The advantage of
dealing with composition-conserving complexes is that all reference
constants involved in their energies and entropies cancel out. This allows
us to directly compare results obtained by different calculation
methods. The complex energies and entropies can be expressed in terms of the
``raw" energies $\epsilon_d$ and entropies $s_d$, $d$ = V$_{\rm Ti}$,
V$_{\rm Al}$, Ti$_{\rm Al}$, Al$_{\rm Ti}$.\cite{MH00} The expressions for
some of the complex energies are given in Table~\ref{pdtial:tab}. Similar
expressions hold for the complex entropies, except that the cohesive energy
$E_0$ should be replaced by the perfect lattice entropy per atom.
Table~\ref{pdtial:tab} summarizes the results of the EAM calculations for
several defect complexes and compares them with the \abinit energies
reported by Woodward \etal\cite{WKY98} The agreement between the two
calculation methods is reasonable. We emphasize again that point defect
properties of \gtial were not included in the potential fit.

\begin{figure}
\includegraphics{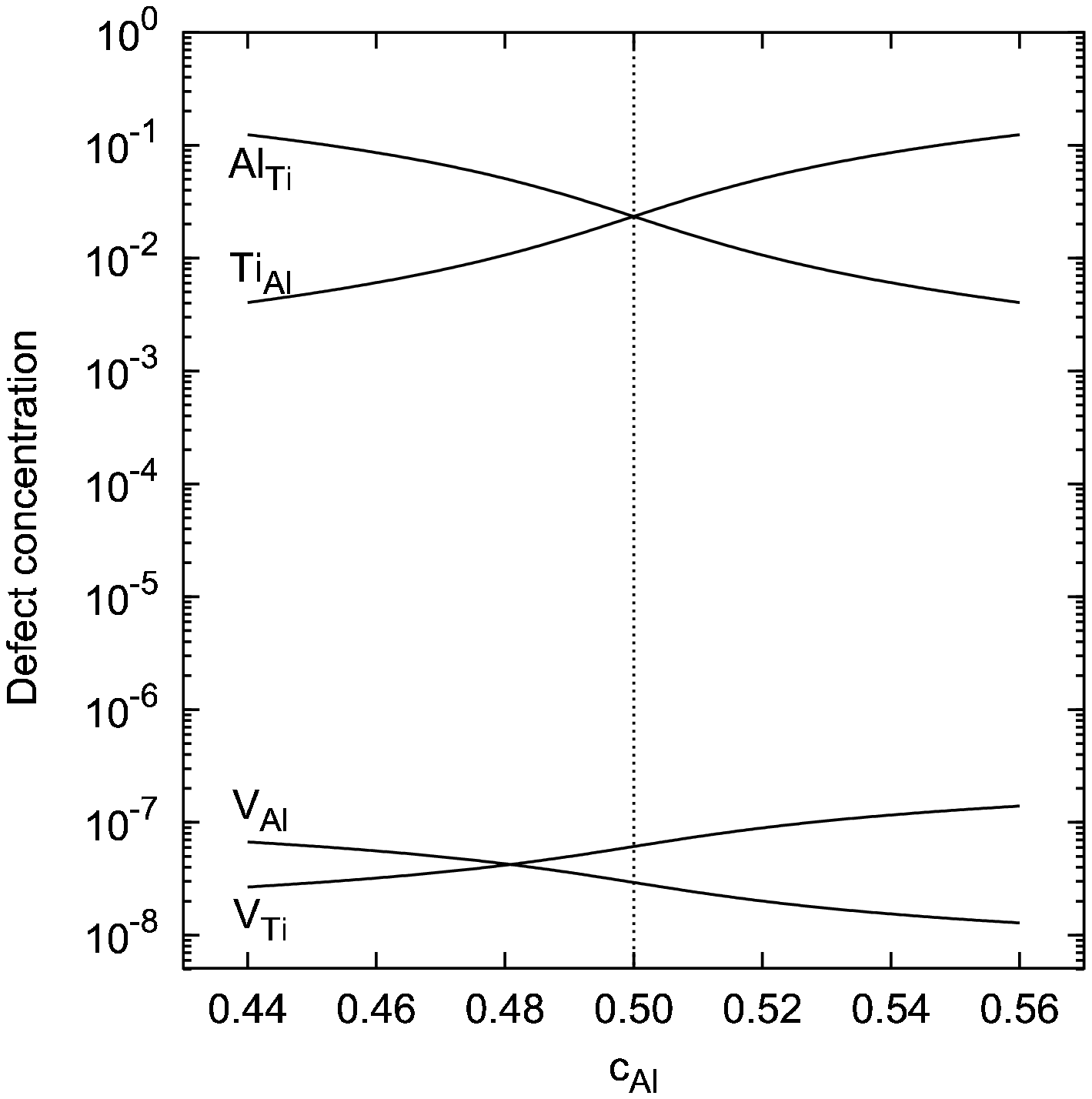}% 
\caption{ Calculated equilibrium concentrations of vacancies and antisites
in TiAl as functions of the alloy composition at 1000 K.}
\label{conc:fig}
\end{figure}

Using the complex energies and entropies, the equilibrium defect
concentrations have been calculated as functions of the bulk composition
around the stoichiometry within the lattice gas model of non-interacting
defects.\cite{MH00,Hagen98,Pasha2000,Fahnle99} Fig.~\ref{conc:fig} shows the
calculation results for $T=1000$ K. We see that all compositions are
strongly dominated by antisite defects. This observation is well consistent
with the experimentally established fact that \gtial is an antisite disorder
compound.\cite{ER54,BWBS94,SY92} The vacancy concentrations are several
orders of magnitude smaller than antisite concentrations. In the
stoichiometric composition, most of the vacancies reside on the Ti
sublattice. All these features have been observed at all temperatures in the
range 800-1200 K.

\section{Summary}

EAM potentials have been developed for Al, \alphati and \gtial by fitting to
a database of experimental data and \abinit calculations. The potentials
have been tested against other experimental and \abinit data not included in
the fitting database. The \abinit structural energies for Ti were calculated
previously,\cite{Mehl02a} while those for Al and Ti-Al compounds have been
generated in this work. All these calculations employed the full-potential
LAPW method within the GGA approximation. Besides serving for the
development of the EAM potentials, the obtained \abinit energies are also
useful as reference data for the Ti-Al system.

The Al potential is fit to the target properties very accurately and has
demonstrated a good performance in the tests. It has certain advantages over
the previously developed potential,\cite{MFMP99} particularly with respect
to the lattice thermal expansion and the pressure-volume relation under
large compressions. The fit of the Ti potential is less successful,
presumably because of the directional component of interatomic bonding that
is not captured by the central-force EAM model. In particular, the potential
underestimates the stacking fault energies on the basal plane. Further
improvements of the potential do not appear to be possible within the
EAM. This potential can be viewed as a supporting potential for the Ti-Al
system, but we also believe that it can be useful in atomistic simulations
in pure Ti where subtle details of atomic interactions may not be critical.
Since the potential is fit reasonably well to the elastic constants, thermal
expansion factors and the vacancy formation energy, it can be employed for
modeling diffusion and creep in large systems that are not accessible by
more accurate, yet slower, \abinit methods.

For the \gtial compound, the potential developed here reproduces reasonably
well the basic lattice properties, planar fault energies, as well as point
defect characteristics. The fit to the elastic constants is better than with
previous potentials. However, the negative Cauchy pressures in \gtial have
not been reproduced by the present nor previous EAM potentials. The planar
fault energies calculated with the potential are in a good agreement with
experiment, but the APB energy is lower than all \abinit values. The fit to
\abinit energies of alternative structures of TiAl enhances the
transferability of the potential to configurations away from
equilibrium. This fact is verified by the good agreement between the EAM and
LAPW energies along the Bain transformation path. The potential also
correctly predicts the equilibrium DO$_{19}$ structure of \ti3al and gives a
fairly good agreement with experiment for the cohesive energy, lattice
parameters, and elastic constants of this compound. The point defect
energies and entropies in \gtial calculated with the potential are in
agreement with the antisite disorder mechanism established for this compound
experimentally. We emphasize that neither the Bain path nor any information
on \ti3al or point-defect properties in \gtial were included in the fitting
database. This success of the proposed potential points to its ability to
describe atomic interactions in the Ti-Al system on a reasonable
quantitative level. The potential should be suitable for large-scale
atomistic simulations of plastic deformation, fracture, diffusion, and other
processes in \gtial. At the same time we acknowledge that more rigorous
models, particularly those including angular-dependent interactions, are
needed for addressing the negative Cauchy pressures, high APB energy, and
other properties of \gtial that lie beyond the capabilities of the EAM.

\begin{acknowledgments}
We would like to thank A.~Suzuki for helpful discussions and assistance with
some of the calculations. We are grateful to M.~J.~Mehl and
D.~A.~Papaconstantopoulos for making the LAPW energies for Ti available to
us prior to publication,\cite{Mehl02a} as well as for helpful comments on
the manuscript. We are also grateful to R.~Pasianot for sending us the
potential of Ref.~\onlinecite{FMP95} in a convenient format, and for very
useful comments on EAM for hcp metals.  This work was supported by the US
Air Force Office of Scientific Research through Grant No.~F49620-01-0025.

\end{acknowledgments}

\def\vr{\vec{r}}
\def\vti{V$_{Ti}$}
\def\val{V$_{\!Al}$}
\def\tial{Ti$_{\!Al}$}
\def\alti{$\epsilon_{{\rm Al}_{\rm Ti}}$}
\def\evti{$\epsilon_{V_{Ti}}$}
\def\eval{$\epsilon_{V_{\!Al}}$}
\def\etial{$\epsilon_{Ti_{\!Al}}$}
\def\ealti{$\epsilon_{Al_{\!Ti}}$}
\def\ez{$\epsilon_0$}
\def\ez{$E_0$}
\def\efi{E$_f^I$}
\def\gsf{$\gamma_{SF}$}
\def\gus{$\gamma_{us}$}
\def\ggb{$\gamma_{gb}$}
\def\gzt{$\gamma_{T}$}
\def\gzs{$\gamma_{s}$}
\def\cxx{C$_{11}$}
\def\cxy{C$_{12}$}
\def\cxz{C$_{13}$}
\def\crxx{C$_{33}$}
\def\cyy{C$_{44}$}
\def\czz{C$_{66}$}

%%%%%%%%%%%%%%%%%%%%%%%%%%%%%%%%%%%%%%%%%%%%%%%%%%%%%%%%%%
% Table: Fitting parameters.   fcc Al
\begin{table}
\caption{
Optimized values of the fitting parameters of the EAM potential for the
Ti-Al system. 
}
\begin{ruledtabular}
\begin{tabular}{llllll}
 \multicolumn{2}{c}{Al}    & \multicolumn{2}{c}{Ti}  &  \multicolumn{2}{c}{TiAl} \\
Parameter & Optimal value  & Parameter & Optimal value & Parameter & Optimal value  \\
\hline
$r_c$ (\AA)    &   6.724884                 & $r_c$ (\AA)  &     5.193995 
& $r_c$ (\AA)     &   5.768489  \\   
$h$  (\AA)      &   3.293585                & $h$ (\AA)    &             0.675729  
& $h$  (\AA)    &   0.619767  \\  
$V_0$ (eV)  &   $-$3.503182 $\times ~ 10^3$   & $V_0$  (eV) &        $-$3.401822 $\times  ~10^{6}$   &
$V_0$ (eV)  &  $-$0.737065  \\ 
$r^{\prime}$(\AA)  & 2.857784               &  $r_1$ (\AA)  &         $-$8.825787  
& $r_{0}$ (\AA)   & 2.845970 \\  
$b_1$          & 8.595076 $\times  ~10^{-2}$ &  $\beta_1$ (1/\AA)   &     5.933482  
& $b_1$  &   5.980610  \\
$b_2$          & 5.012407 $\times  ~10^{-2}$ & $V_0^{\prime}$ (eV)   &  0.161862   
& $b_2$  &   5.902127  \\ 
$\delta$(eV)   & 3.750298 $\times  ~10^3$    & $r_1^{\prime}$(\AA)  &  3.142920     
&   $\delta$ (eV) & 0.078646  \\  
$y$               & 2.008047 $\times  ~10^1$ & $\beta_2$ (1/\AA)   &     2.183169     
&  $s_{Al}$  &   0.951039 \\  
$\gamma$(1/\AA)   &  4.279852                & $\delta$ &        $-$0.601156  $\times  ~10^{-1}$ 
& $g_{Ti}$   (eV) &4.839906 \\ 
$B_0$(\AA)          &  1.192727 $\times  ~10^5$    &  $A$    &            3.656883  $\times  ~10^{2}$   
&   $g_{Al}$   (eV) &1.281479 \\  
$C_0$    (1/\AA$^3$) &  8.60297  $\times  ~10^{-2}$ &  $r_0$ (\AA)  &         $-$1.169053  $\times  ~10^{1}$     
&  & \\
$r_0$(\AA)     &   0.5275494                     &  $r_0^{\prime}$(\AA) &  $-$2.596543  $\times  ~10^{2}$  
& &    \\
$\beta$ & 0.00489   &  $\alpha_1$ (1/\AA)   &    0.3969775 $\times  ~10^{-1}$   
& &    \\
&  &  $\alpha_2$  (1/\AA)  &    5.344506  $\times  ~10^{2}$   
& &    \\
&  &  $B_1$         &    1.549707  
& &    \\
&  &  $B_2$         &   $-$0.4471131 
& &    \\
&  &  $B_3$         &    0.8594003 $\times  ~10^{-1}$    
& &    \\
\end{tabular}
\end{ruledtabular}
\label{tab:param}
\end{table}

% Table: Equilibrium properties.   Al
\begin{table}
\caption{
 Properties of Al calculated using the present EAM potential and the MFMP
potential\cite{MFMP99} 
in comparison with experimental data.}
\begin{ruledtabular}
\begin{tabular}{lcll}
 Property               & Experiment      &  EAM       &  MFMP   \\
\hline
Lattice properties:     &                 &                 &          \\
 a$_0$ (\AA)            & 4.05\footnotemark[1]       &   4.05         &  4.05     \\
 E$_0$ (eV/atom)        & 3.36\footnotemark[2]      &   3.36         &  3.36     \\
 B (GPa)       & 79\footnotemark[3]       &   79         &  79       \\
 \cxx (GPa)     & 114\footnotemark[3]       &   116.8        &  113.8        \\
 \cxy (GPa)     & 61.9\footnotemark[3]      &   60.1        &  61.6        \\
 \cyy (GPa)     & 31.6\footnotemark[3]      &   31.7        &
31.6 \\
Vacancy:    &                   &                 &          \\
 E$_v^f$(eV)           & 0.68\footnotemark[4]  &  0.71     &   0.68    \\
 E$_v^m$(eV)           & 0.65\footnotemark[5]  &  0.65     &   0.64    \\
$\Omega_v^f/\Omega_0$ &  0.62\footnotemark[6]         &  0.59     & 0.51     \\

Planar defects:    &                   &                 &          \\
\gsf(\mjm2)     & 166\footnotemark[7],120-144\footnotemark[8]   &    115   & 146   \\
\gus(\mjm2)     &                               &    151   & 168   \\
\gzt(\mjm2)     &  76\footnotemark[7]           &    63   & 76   \\

Surface: &                   &                 &          \\
\gzs(110)(\mjm2)   & 980\footnotemark[9],~1140\footnotemark[10],~1160\footnotemark[11] &  792      &    1006   \\
\gzs(110)(\mjm2)   & 980\footnotemark[9],~1140\footnotemark[10],~1160\footnotemark[11] &  607      &    943   \\
\gzs(111)(\mjm2)   & 980\footnotemark[9],~1140\footnotemark[10],~1160\footnotemark[11] &  601      &    870   \\
\end{tabular}
\end{ruledtabular}
\label{aleq:tab}
\footnotetext[1]{Ref.~\onlinecite{Kittel}}
\footnotetext[2]{Ref.~\onlinecite{CRC}}
\footnotetext[3]{Ref.~\onlinecite{SW77}}
\footnotetext[4]{Ref.~\onlinecite{SGSS}}
\footnotetext[5]{Ref.~\onlinecite{Ball78}}
\footnotetext[6]{Ref.~\onlinecite{Emrick69}}
\footnotetext[7]{Ref.~\onlinecite{Murr75}}
\footnotetext[8]{Ref.~\onlinecite{Raut82} and \onlinecite{WP71}} 
\footnotetext[9]{Average orientation, Ref.~\onlinecite{Murr75}} 
\footnotetext[10]{Average orientation, Ref.~\onlinecite{TM77}}
\footnotetext[11]{Average orientation, Ref.~\onlinecite{BBMMN88}}
\end{table}

% Table: Structural Energy diff. for Al.
\begin{table}
\caption{Comparison of the energies (eV/atom) of selected structures of Al
calculated by the LAPW method, with the present EAM potential, and with the
MFMP potential.\protect\cite{MFMP99} The energies are given relative to the
energy of the equilibrium fcc structure.}
% revised version for resubmission: Raja
\begin{ruledtabular}
\begin{tabular}{lccc}
Structure   &LAPW         &  EAM    &   MFMP \\
\hline
hcp        &  0.04   & 0.03    & 0.03   \\
bcc        &  0.09   & 0.09  & 0.011   \\
L1$_0$     &  0.27  & 0.33  & 0.30   \\
sc         &  0.36  & 0.30  & 0.40   \\
diamond    &  0.75  & 0.88  & 0.89   \\
\end{tabular}
\end{ruledtabular}
\label{edfal:tab}
\end{table}

% Thermal expansion for Al.
\begin{table}
\caption{ The linear thermal expansion factor (in \%) of Al computed using
the present EAM potential in comparison with experimental data at selected
temperatures. QHA: quasiharmonic approximation; MC: Monte Carlo method.}
\begin{ruledtabular}
\begin{tabular}{cccccc}
 T (K) & Experiment\footnote{Ref.~\onlinecite{TKTD75}} &
\multicolumn{2}{c}{EAM} \\ 
& & QHA & MC & \\ 
\hline 
293 & 0.418 & 0.277 & 0.489 & \\ 
500 & 0.932 & 0.663 & 0.872 & \\ 
700 & 1.502 & 1.016 & 1.332 & \\ 
900 & 2.182 & 1.419 & 1.916 & \\
\end{tabular}
\end{ruledtabular}
\label{teal:tab}
\end{table}

% Table: EAM properties   Ti
\begin{table}
\caption{Properties of Ti predicted by the present EAM potential and the
FMP\cite{FMP95} potential in comparison with experimental data. }
\begin{ruledtabular}
\begin{tabular}{lcrr}
             & Experiment & EAM  & FMP\footnote{Ref.~\onlinecite{FMP95}}  \\
\hline
a$_0$ (\AA)  &  2.951\footnotemark[1]       & 2.951    & 2.951 \\
c/a      &  1.588\footnotemark[1]       & 1.585    & 1.588  \\
E$_0$ (eV/atom)   &  4.850\footnotemark[2]       & 4.850    & 4.850 \\
\cxx(GPa)&   176\footnotemark[3]    & 178    & 189  \\
\cxy(GPa)&   87\footnotemark[3]    & 74    & 74  \\
\cxz(GPa)&   68\footnotemark[3]    & 77    & 68  \\
\crxx(GPa)&  190\footnotemark[3]    & 191    & 188  \\
\cyy(GPa)&   51\footnotemark[3]    & 51    & 50  \\
 \evf (eV)      & 1.55\footnotemark[4]       &    1.83  &  1.51  \\
 \evm(basal) (eV)   &            &   0.80      &  0.51   \\
 \evm(nonbasal) (eV)&            &   0.83      &  0.48   \\
 $Q$ (eV)              &   3.14\footnotemark[6]     &   2.62      &  2.02    \\
 $\gamma_{I_1}$ (\mjm2)  &            &  31           &   31        \\
 $\gamma_{I_2}$ (\mjm2)  & 290\footnotemark[7],300\footnotemark[8]           &  56           &   57        \\
 $\gamma_{E}$ (\mjm2)  &      &  82           &   84        \\
 $\gamma_{s}$ (0001) (\mjm2)  & 2100\footnotemark[9], 1920\footnotemark[10]     & 1725          &  1439       \\
\end{tabular}
\end{ruledtabular}
\label{tieq:tab}
\footnotetext[1]{Ref.~\onlinecite{Pear87}}
\footnotetext[2]{Ref.~\onlinecite{Kittel}}
\footnotetext[3]{Ref.~\onlinecite{SW77}}
\footnotetext[4]{Ref.~\onlinecite{Shestopal66}}
\footnotetext[6]{Ref.~\onlinecite{KHFM97}}
\footnotetext[7]{Ref.~\onlinecite{Legrand84}}
\footnotetext[8]{Ref.~\onlinecite{Patridge67}}
\footnotetext[9]{Average orientation, Ref.~\onlinecite{BBMMN88}}
\footnotetext[10]{Average orientation, Ref.~\onlinecite{TM77}}
\end{table}

% Table: Structural Energy diff. for hcp titanium
\begin{table}
\caption{ Energies (eV/atom) of selected structures of Ti obtained with the
present EAM potential and by the LAPW/GGA-PW91 method.\protect\cite{Mehl02a}
All energies are given relative to the energy of the experimentally observed
hcp structure.}
\begin{ruledtabular}
\begin{tabular}{llll}
Structure   &   LAPW/GGA-PW91    &  EAM    &   FMP \\
\hline
fcc        &  ~~0.012 & 0.011 & 0.012 \\
bcc        &  ~~0.067  & 0.03  & 0.02  \\
sc         &  ~~0.77  & 0.54  & 0.27  \\
omega      & $-$0.06  & 0.094  & 0.064   \\
\end{tabular}
\end{ruledtabular}
\label{edfti:tab}
\end{table}

% Thermal expansion for the hcp Titanium.
\begin{table}
\caption{The linear thermal expansion factor (in \%) of Ti calculated with the
present EAM potential in comparison with experimental data at selected
temperatures.  QHA: quasiharmonic approximation; MC: Monte Carlo
method.}
\begin{ruledtabular}
\begin{tabular}{rllccl}
 T (K)    & Experiment\footnote{Ref.~\onlinecite{TKTD75}}  &      \multicolumn{2}{c}{EAM}         &   \\
               &        &      QHA   & MC &   \\
\hline
 293      & 0.15       & 0.16       &  0.25        \\
 500      & 0.35       & 0.38       &  0.44       \\
 700      & 0.55       & 0.62       &  0.63       \\
1000      & 0.89       & 1.00       &  0.72        \\
\end{tabular}
\end{ruledtabular}
\label{teti:tab}
\end{table}

% Table: TiAl EAM properties
\begin{table}
\caption{
Equilibrium lattice constant,  $c/a$ ratio, cohesive energy,  and 
elastic constants of \gtial calculated with the present EAM potential in
comparison with other potentials\cite{SRD97,Farkas94} and experimental data.}
\begin{ruledtabular}
\begin{tabular}{lcrrrr}
             & Experiment &  EAM &
Ref.~\protect\onlinecite{Farkas94} & Ref.~\protect\onlinecite{SRD97}  \\
\hline
$a_0$ (\AA)  &  3.997\footnotemark[1]     & 3.998 & 3.951 & 4.033   \\
$c/a_0$      &  1.02\footnotemark[1]          & 1.047 & 1.018 & 0.991   \\
$E_0$ (eV/atom)   &  4.51\footnotemark[2]     &  4.509  & 4.396 & 4.870  \\
\cxx(GPa)    &  186\footnotemark[3], 183\footnotemark[4]     & 195 & 222 & 202
\\   
\cxy(GPa)    &   72\footnotemark[3], 74.1\footnotemark[4]     & 107 & 100 &
95  \\ 
\cxz(GPa)    &   74\footnotemark[3], 74.4\footnotemark[4]     & 113 & 162 &
124    \\ 
\crxx(GPa)    &  176\footnotemark[3], 178\footnotemark[4]     & 213 & 310 & 237
\\  
\cyy(GPa)    &  101\footnotemark[3], 105\footnotemark[4]     & 92 & 139 & 83
\\ 
\czz(GPa)    &  77\footnotemark[3],  78.4\footnotemark[4]     & 84  & 76 & 54
\\ 
\end{tabular}
\end{ruledtabular}
\label{eqtial:tab}
\footnotetext[1]{ Ref.~\onlinecite{Pear87}}
\footnotetext[2]{ Ref.~\onlinecite{HOAK63}}
\footnotetext[3]{ Ref.~\onlinecite{HSMW95}}
\footnotetext[4]{ Ref.~\onlinecite{TK96}}
\end{table}

% Planar defects (TiAl) 
\begin{table}
\caption{ The energies of the superlattice intrinsic stacking fault (SISF),
antiphase boundary (APB), complex stacking fault (CSF), and the (100) and
(111) surfaces in \gtial calculated with the present EAM
potential. Available experimental and \abinit data are included
for comparison. All energies are expressed in \mjm2.}
\begin{ruledtabular}
\begin{tabular}{lccc}
                     &  Experiment    & {\it ab-initio} &  EAM     \\ 
\hline 
 SISF(111)           &  140\footnotemark[1]     
 & 90\footnotemark[2], 110\footnotemark[1], 123\footnotemark[3],
172\footnotemark[4]    &  173       \\ 
 CSF(111)            &           
& 280\footnotemark[1], 294\footnotemark[3], 363\footnotemark[4]            &
299       \\ 
 APB(111)            &  250\footnotemark[1] 
 &  510\footnotemark[2], 667\footnotemark[4], 670\footnotemark[1],  672\footnotemark[3]   &   266       \\
surface  (100) &           &                                    &  1177      \\
surface  (110) &           &                                    &  1445       \\
\end{tabular}
\end{ruledtabular}
\label{sftial:tab}
\footnotetext[1]{Ref.~\onlinecite{WMR92}}
\footnotetext[2]{Refs.~\onlinecite{FY90,FY93}}
\footnotetext[3]{Ref.~\onlinecite{WM96}}
\footnotetext[4]{Ref.~\onlinecite{EF98}}

\end{table}

\begin{table*}
\caption{Formation energies, $\Delta H$ (eV/atom), of different compounds of
the Ti-Al system calculated with the present EAM potential and by the LAPW
method in comparison with literature data.}

\label{eform:tab}
\begin{ruledtabular}
\begin{tabular}{lcllcl}
     \multicolumn{3}{c}{TiAl}   &   \multicolumn{3}{c}{Ti$_3$Al}  \\
\hline
 Structure &  $\Delta H$        &   Method % TiAl
& Structure& $\Delta H$        &   Method \\
\hline
 L1$_0$     &   $-$0.404       &           EAM\footnotemark[1]     % TiAl
  & DO$_{19}$  &   $-$0.289  &         EAM\footnotemark[1] \\
 L1$_0$     &   $-$0.43        &        LAPW/GGA\footnotemark[1]     % TiAl
   & DO$_{19}$  &   $-$0.318  &    LAPW/GGA\footnotemark[1] \\
 L1$_0$     & $-$(0.37-0.39)   &        Experiment\footnotemark[2]    % TiAl
   & DO$_{19}$  &   $-$0.25,$-$0.26  &    Experiment\footnotemark[6] \\
 L1$_0$     &   $-$0.38        &        Experiment\footnotemark[3]     % TiAl
 & DO$_{19}$  &   $-$0.28   &         FLASTO/LDA\footnotemark[4] \\
 L1$_0$     &   $-$0.41        &        FLASTO/LDA\footnotemark[4]     % TiAl
 & DO$_{19}$  &   $-$0.29, $-$0.28   &  LMTO/LDA\footnotemark[7],  FLAPW/LDA\footnotemark[7] \\
 L1$_0$     &   $-$0.44        &        FLMTO/LDA\footnotemark[5]     % TiAl
  & DO$_{19}$  &   $-$0.28          &  FLAPW/LDA\footnotemark[8]  \\
 B$_2$      &   $-$0.27        &        EAM\footnotemark[1]     % TiAl
 & L$_{12}$   &   $-$0.288  &         EAM\footnotemark[1] \\
 B$_2$      &   $-$0.29        &        LAPW/GGA\footnotemark[1]     % TiAl
&  L$_{12}$  &   $-$0.30       &  LAPW/GGA\footnotemark[1]\\
 B$_2$      &   $-$0.26        &        FLASTO/LDA\footnotemark[4]     % TiAl
& L$_{12}$   &   $-$0.28  &          FLASTO/LDA\footnotemark[4] \\
 B$_1$      &    0.13        &        EAM\footnotemark[1]     % TiAl
&  L$_{12}$  &   $-$0.29, $-$0.27   &  LMTO/LDA\footnotemark[7], FLAPW/LDA\footnotemark[7] \\
 B$_1$      &    0.24        &        LAPW/GGA\footnotemark[1]     % TiAl
& DO$_{11}$  &   $-$0.03   &         EAM\footnotemark[1] \\
 L1$_1$     &   $-$0.30        &        EAM\footnotemark[1]     % TiAl
 & DO$_{22}$  &   $-$0.28   &         EAM\footnotemark[1] \\
 B32        &   $-$0.32        &        EAM\footnotemark[1]     % TiAl
 & DO$_{22}$  &   $-$0.27, $-$0.25   &  LMTO/LDA\footnotemark[7], FLAPW/LDA\footnotemark[7] \\
 ``40''     &   $-$0.37        &        EAM\footnotemark[1]     % TiAl
& DO$_3$     &   $-$0.23   &         EAM\footnotemark[1] \\
\hline
\multicolumn{3}{c}{Al$_3$Ti}   &  & \\
 Structure &  $\Delta H$        &   Method % Al3Ti
 &  & & \\
% The following data is for Al3Ti.
DO$_{22}$  &   $-$0.29  &         EAM\footnotemark[1] % Al3Ti
 &  & & \\
              DO$_{22}$  &   $-$0.41   &         FLASTO/LDA\footnotemark[4] % Al3Ti
 &  & & \\
              DO$_{22}$  &   $-$0.42   &         LMTO/LDA\footnotemark[9] % Al3Ti
 &  & & \\
               L$_{12}$  &   $-$0.30  &         EAM\footnotemark[1] % Al3Ti
 &  & & \\
              DO$_3$     &   $-$0.20   &         EAM\footnotemark[1] % Al3Ti
 &  & & \\
              DO$_{19}$  &   $-$0.29  &         EAM\footnotemark[1] % Al3Ti
 &  & & \\
\end{tabular}
\end{ruledtabular}
\footnotetext[1]{Present work}
\footnotetext[2]{Ref.~\onlinecite{BBMMN88,Desai87}}
\footnotetext[3]{Ref.~\onlinecite{BB92}}
\footnotetext[4]{Ref.~\onlinecite{WW98}}
\footnotetext[5]{Ref.~\onlinecite{AFSSM92}}
\footnotetext[6]{Refs.~\onlinecite{BBMMN88,HDHGK73}}
\footnotetext[7]{Ref.~\onlinecite{HWGFOX91}}
\footnotetext[8]{Ref.~\onlinecite{FZY95}}
\footnotetext[9]{Ref.~\onlinecite{HWFOX90}}
\end{table*}

%%%%%%%%%%%%%%%%%%%%%%%%%%%%%%%%%%%%%%%%%%%%%%%%%%%%%%%%%%%%%%%%%%%%%%%%%%%%%%%%%%%%%%%%

\begin{table}
\caption{ Equilibrium properties of Ti$_3$Al predicted by the present EAM
potential. {\em Ab initio}\ and experimental data are included for
comparison.}
\begin{ruledtabular}
\begin{tabular}{llll}
Property  &  Experiment                  &  EAM    & \abinit       \\
\hline
a$_0$ (\AA)    &  5.77\footnotemark[1]                &  5.784  &  5.614\footnotemark[5]    \\
c/a        &  0.8007\footnotemark[1]              &  0.821  &  0.831\footnotemark[5]    \\
E$_0$ (eV/atom)     &  4.78\footnotemark[2]                &  4.766  &      \\
\cxx (GPa)     &  176.2\footnotemark[3], 175\footnotemark[4]      &  180.5  &    221\footnotemark[6] \\
\cxy (GPa)     &  87.8\footnotemark[3],  88.7\footnotemark[4]     &  74.4   &    71\footnotemark[6] \\
\cxz (GPa)     &  61.2\footnotemark[3],  62.3\footnotemark[4]     &  70.3   &    85\footnotemark[6] \\
\crxx (GPa)    &  218.7\footnotemark[3], 220\footnotemark[4]      &  222.9  &    238\footnotemark[6] \\
\cyy  (GPa)    &  62.4\footnotemark[3],  62.2\footnotemark[4]     &  46.6   &    69\footnotemark[6] \\
\end{tabular}
\end{ruledtabular}
\label{ti3al:tab}
 \footnotetext[1]{Ref.~\onlinecite{Pear87}}
 \footnotetext[2]{Ref.~\onlinecite{HDHGK73}}
 \footnotetext[3]{Ref.~\onlinecite{HSMW95}}
 \footnotetext[4]{Ref.~\onlinecite{TOIMYK96}}
 \footnotetext[5]{Ref.~\onlinecite{WW98}}
 \footnotetext[6]{Ref.~\onlinecite{FZY95}}
\end{table}

% Thermal expansion for TiAl.
\begin{table}
\caption{The linear thermal expansion factor (in \%) of \gtial calculated
with the present EAM potential.  QHA: quasiharmonic approximation; MC:
Monte Carlo method.}
\begin{ruledtabular}
\begin{tabular}{ccc}
 T (K)    &       \multicolumn{2}{c}{EAM} \\
          &              QHA   & MC    \\
\hline
 400              &    0.36    & 0.53       \\
 600              &    0.66    & 0.85       \\
 800              &    1.03    & 1.20       \\
 1000             &    1.54    & 1.58       \\
\end{tabular}
\end{ruledtabular}
\label{te:tab}
\end{table}

\begin{table*}
\caption{
 Energies (in eV) and entropies (normalized to k$_B$) of point defect complexes in TiAl computed
using the present EAM potential. \Abinit results are included for comparison.}
\begin{ruledtabular}
\begin{tabular}{llccc}
Complex   &       Equation         & {\it ab-initio}\footnote{Ref.~\onlinecite{WKY98}}
& \multicolumn{2}{c}{EAM} \\
          &                        &  Energy    &  Energy   & Entropy (k$_B$)   \\
\hline
Divacancy &  \evti+\eval+2\ez      &   3.582        &  3.168    &   2.804    \\
Exchange  &  \ealti+\etial         &   1.204        &  0.765    &   1.420    \\
Triple-Ti & 2\evti+2\ez+\etial     &  3.775         &  3.525    &   3.952    \\
Triple-Al & 2\eval+2\ez+\ealti    &  4.593         &  3.576    &   3.075    \\
Interbranch Ti & 2\eval-\etial+2\ez & 3.389         &  2.810    &   1.656    \\
Interbranch Al & \ealti-2\evti-2\ez & $-$2.571        & $-$2.759    &  $-$2.532    \\
\end{tabular}
\end{ruledtabular}
\label{pdtial:tab}
\end{table*}

\begin{thebibliography}{113}
\bibitem[{Far()}]{Farkas-elast}
\bibinfo{note}{The version of the potential available to us does not reproduce
  the elastic constant reported in Table 2 of
  Ref.~\protect\onlinecite{Farkas94}, whereas all other properties are
  reproduced exactly.}
\end{thebibliography}
\end{document}